\documentclass[aps,prb,twocolumn,groupedaddress,showpacs]{revtex4}

\usepackage{longtable}
\usepackage{graphicx}
\usepackage{dcolumn}
\usepackage{bm}
\usepackage{amsmath}
\usepackage[psamsfonts]{amssymb}

\newcommand{\STO}{SrTiO$_3$}
\newcommand{\LAO}{LaAlO$_3$}
\begin{document}

\title{Anomalous response to gate voltage application in mesoscopic \LAO/\STO  devices}


\author{D. Rakhmilevitch}
\affiliation{Raymond and Beverly Sackler School of Physics and
Astronomy, Tel-Aviv University, Tel Aviv, 69978, Israel}
\author{I. Neder}
\affiliation{Raymond and Beverly Sackler School of Physics
and Astronomy, Tel-Aviv University, Tel Aviv, 69978,
Israel}
\author{M. Ben Shalom}
\affiliation{Raymond and Beverly Sackler School of Physics and
Astronomy, Tel-Aviv University, Tel Aviv, 69978, Israel}
\author{A. Tsukernik}
\affiliation{Raymond and Beverly Sackler School of Physics
and Astronomy, Tel-Aviv University, Tel Aviv, 69978,
Israel}
\author{M. Karpovski}
\affiliation{Raymond and Beverly Sackler School of Physics
and Astronomy, Tel-Aviv University, Tel Aviv, 69978,
Israel}
\author{Y. Dagan}
\affiliation{Raymond and Beverly
Sackler School of Physics and Astronomy, Tel-Aviv University, Tel
Aviv, 69978, Israel}
\author{A. Palevski}
\affiliation{Raymond and Beverly Sackler School of Physics and
Astronomy, Tel-Aviv University, Tel Aviv, 69978, Israel}

\date{\today}
\begin{abstract}
We report on resistivity and Hall measurements performed on a series
of narrow mesa devices fabricated from $LaAlO_3/SrTiO_3$ single interface heterostructure with a bridge width range of 1.5-10 microns. Upon applying back-gate voltage of the order of a few Volts, a strong increase in the sample resistance (up to factor of  35) is observed, suggesting a relatively large capacitance between the Hall-bar and the gate. The high value of this capacitance is due to the device geometry, and can be explained within an electrostatic model using the Thomas Fermi approximation.
The Hall coefficient is sometimes a  non-monotonic function of the gate voltage. This behavior is inconsistent with a single conduction band model. We show that a theoretical two-band model is consistent with this transport behavior, and indicates a metal to insulator transition in at least one of these bands.

\end{abstract}
\maketitle
Interfaces between strongly correlated oxides exhibit a variety of physical phenomena, with properties which can be very different from their constituent compounds. The ability to modulate those properties using electric field opens the possibility for new, oxide based, electronics\cite{Oxide_electronics}. A widely studied example for such an interface is the two dimensional electron gas (2DEG) formed at the interface between $LaAlO_3$ (LAO) and $SrTiO_3$ (STO)\cite{OhtomoHwang}. Extensive studies have shown that, under appropriate growing conditions, a critical thickness of four unit cells of LAO is required for the formation of a 2DEG with a superconducting ground state\cite{Thiel,ReyrenSC}. The origin\cite{PhysRevLett.98.196802,Nakagawa_no_O_defects,popovictheoryfortwodeg,pentcheva:035112,pentchevaPicketPRL,kalabukhovOVac,yoshimatsuorigin} and dimensionality\cite{HerranzPRL,Bernhard_elipsometry,Basletic_thickness} of the charge carriers are still under debate.
\par
Further research demonstrated that several of the 2DEG properties, namely the conductivity\cite{Thiel}, spin orbit coupling\cite{BenSHalomPRL,CavigliaSpinorbit} and the transition temperature to the superconducting\cite{CavigiliaGating} state, can be modified by an electric field. However, in order to manipulate the properties of the 2DEG in macroscopic structures, high voltages of  tens to hundreds of volts were required\cite{Bell2009,Joshua2012}. Such voltages are far beyond the voltages used today in the semiconductor industry and therefore restrict the applicability of devices based on the properties of the 2DEG formed in the interface of LAO/STO.
\par
Recently several theoretical\cite{popovictheoryfortwodeg} and experimental\cite{Copie2DEG,Quantumos,SdHCaviglia,Lerer2011} studies showed evidence of multiple types of carriers generated at the LAO/STO interface. According to observations, most of the carriers have low mobility and only a fraction of the carriers contribute to transport phenomena requiring high mobility such as Shubnikov-de Haas oscillations. Moreover it was long suspected that Hall effect measurements do not provide a real estimate of the carriers and that other methods are needed\cite{multiplecarrier}. Previously we have shown that the analysis of phase coherent transport in mesoscopic structures of LAO/STO interface also indicates the existence of multiple bands\cite{UCFLAOSTO}.
\par
Very recently it has been suggested that the 2DEG at LAO/STO interface undergoes a metal-insulator phase transition at some critical electron density \cite{Manhart_2011,Joshua2012}. The suggested model is based on, an experimentaly observed, rapid drop in conductivity within a narrow range  of back gate voltages.
\par
\begin{figure}
\includegraphics[scale=0.5]{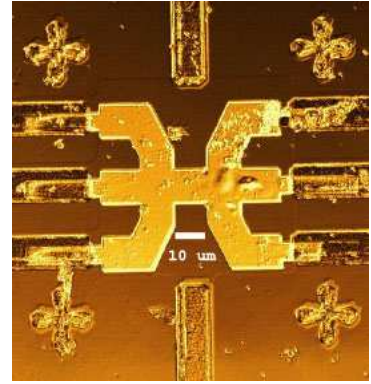}
\caption {Confocal microscope image of measured region in a typical device, before the stage of metal removal. Eight unit cells of \LAO~ were grown on a \STO~substrate, terminated by an atomically flat $TiO_2$ layer by pulsed laser deposition (growth parameters were described elsewhere\cite{benshalom}). Hall-bar geometry devices were patterned on the sample using combination of photolithography and electron-beam lithography. The LAO was dry-etched using Reactive-Ion-Etching (RIE) by Ar ions.\cite{UCFLAOSTO} \label{sample}}
\end{figure}
This paper presents the results of transport measurements performed on narrow mesa devices of LAO/STO. Three key findings are reported: first, all the devices show a strong increase in their resistivity upon applying negative gate voltage of only a few volts. Second, we show that the resistance vs. gate voltage characteristics of the various devices are all self-similar using the capacitance per unit area as a scaling parameter. The capacitance is observed to be inversely proportional to the width of the Hall bar. Finally, the Hall coefficient exhibits a non-monotonic dependence on gate voltage, which strongly suggests contributions from more than one conducting band.
\par
A picture of a typical device, taken with a confocal microscope, is presented in Fig.\ref{sample}. Several Hall bar geometry devices were fabricated, with widths varying from 1.5 $\mu$m to 10 $\mu$m and a constant aspect ratio. A thin gold layer was  evaporated on the bottom of each sample and was used as a back gate. 
\par
\begin{figure}
\begin{tabular}{|c|c|}
\hline 
\includegraphics[scale=0.36]{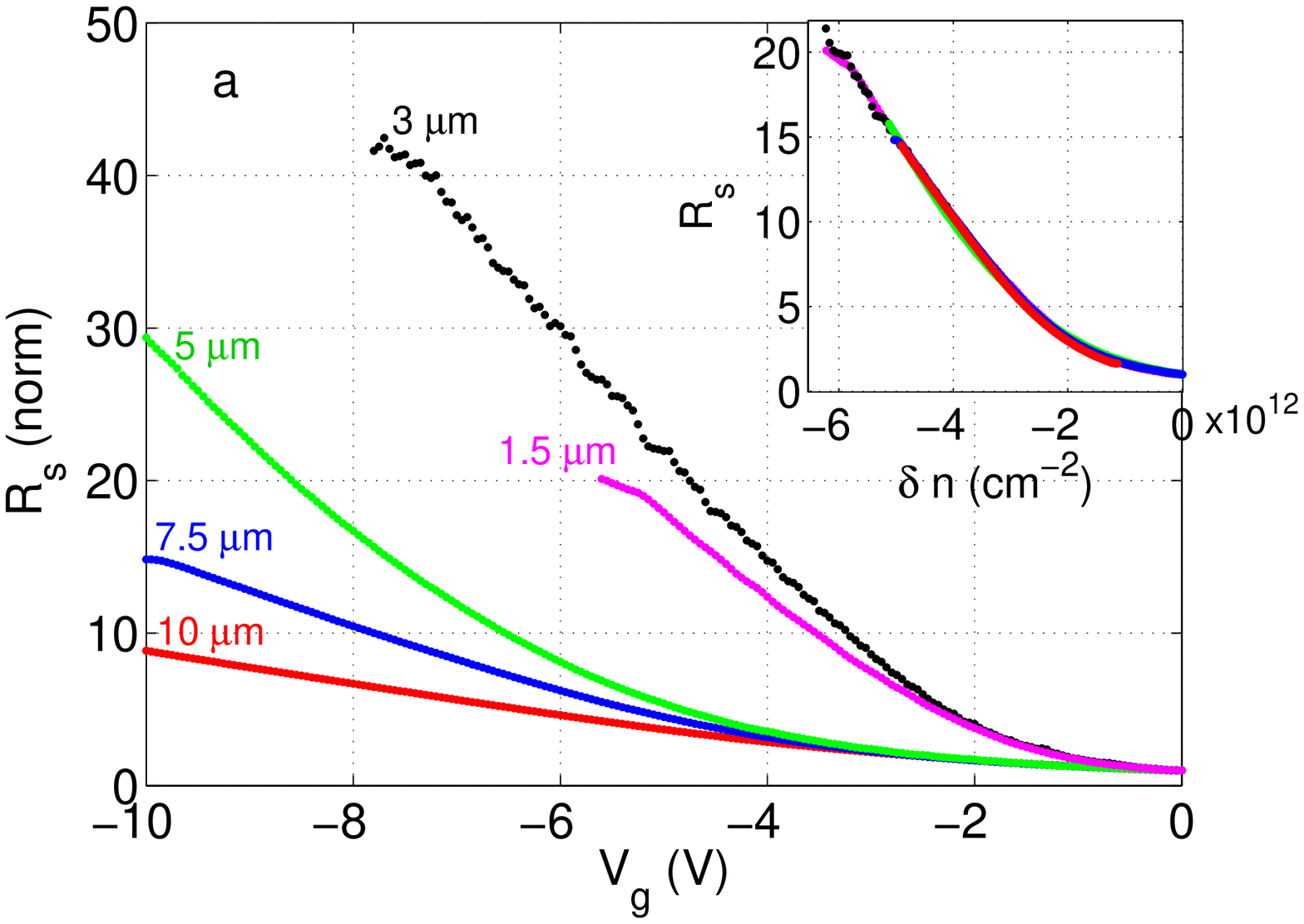}
\tabularnewline
\includegraphics[scale=0.36]{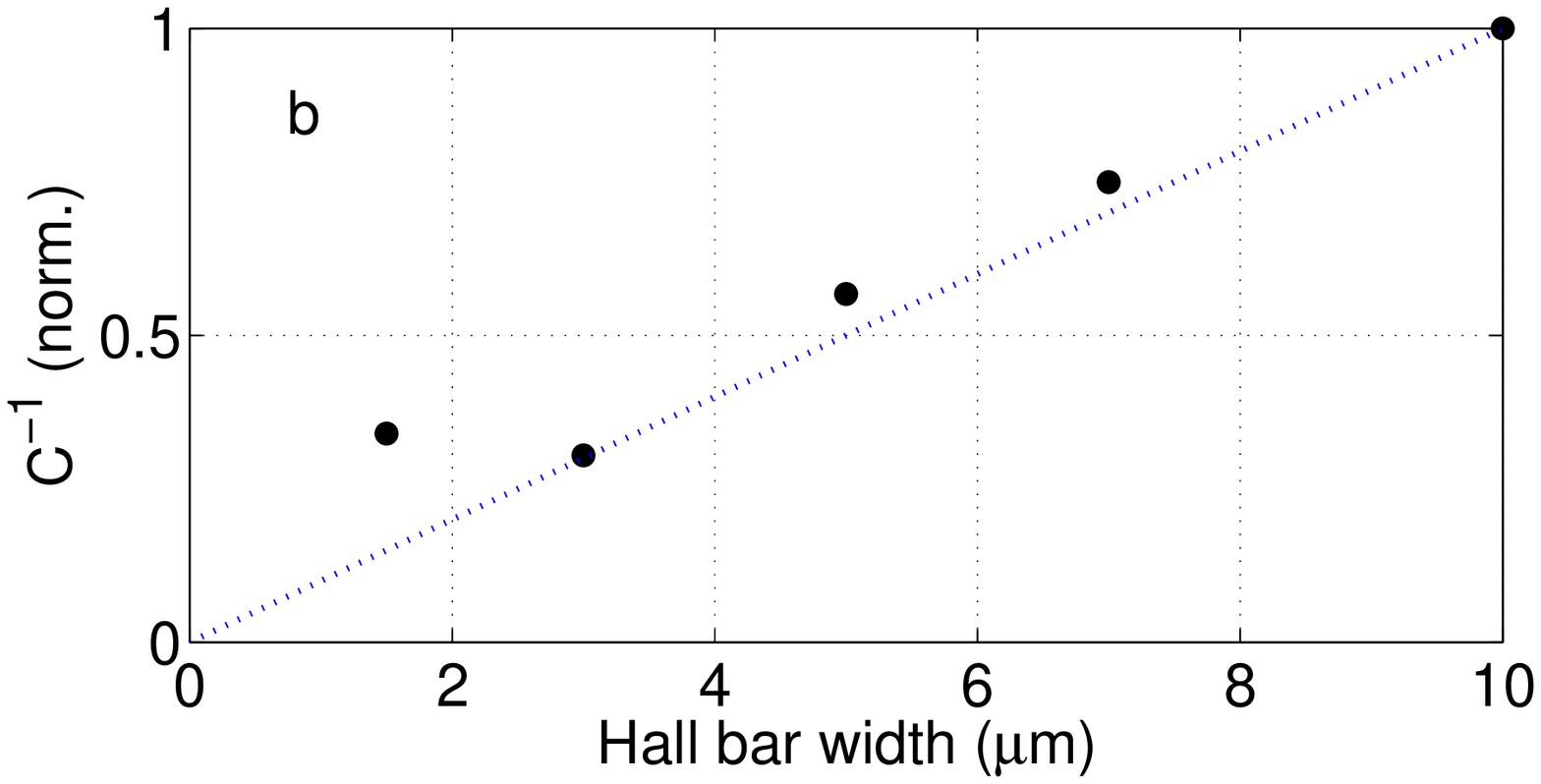} \tabularnewline
\hline 
\end{tabular}
\caption {(color online) (a) Normalized conductance vs.~gate voltage ($V_g$), measured at T=4.2 K. 	The conductance is normalized to its value at $V_g=0$. The different colors correspond to measurements performed on Hall-bars with  different widths, which are indicated near each curve. Inset: The result of shifting and rescaling the x axis of the different curves in Fig. 2a. The $x$-axis is the estimated change in the total electron density due to the change in the gate voltage, given the calculated  geometrical capacitance of the  $10\,\mu m$ device of $1.1\times10^{12}\,el/cm^2/V\pm15\%$ (see text). (b) The inverse of fitted capacitance values for the various Hall-bars which produced the curves in the inset, normalized by the value for the $10 \mu m$ Hall-bar.\label{Gmeas}}
\end{figure}
\begin{figure}
\begin{tabular}{|l|l|}
\hline 
 \includegraphics[scale=0.35]{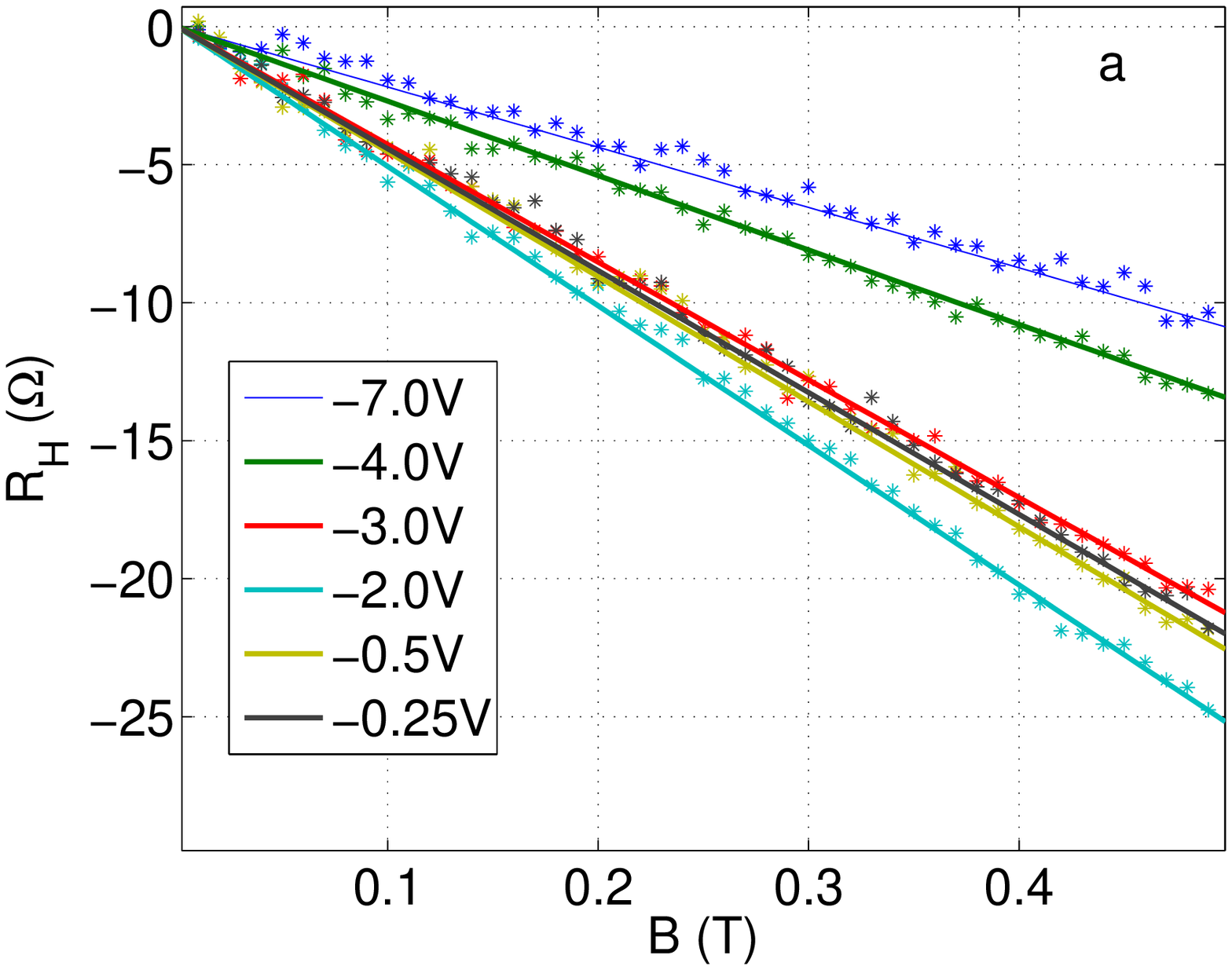}\tabularnewline
\includegraphics[scale=0.35]{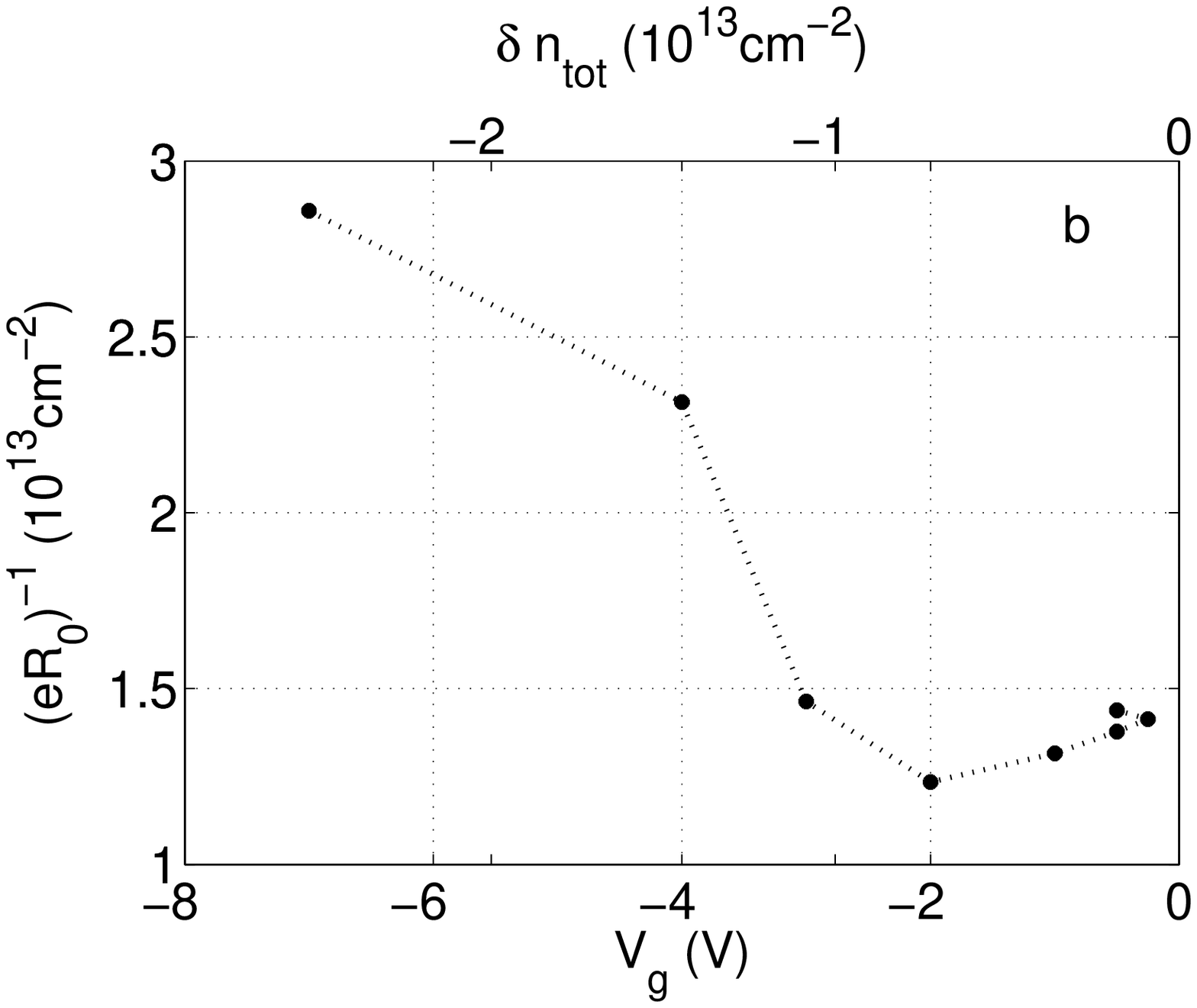} \tabularnewline
\hline 
\end{tabular}\caption {Hall measurements results for the 3 $\mu m$ wide device, measured at 4.2 K. (a) The Hall resistivity curves, $R_H$ as a function of magnetic field, at certain values of the gate voltage $V_g$. The curves are linear, but their slopes is non-monotonic in $V_g$.	 (b) The inverse of the slope $R_0=-\frac{dR_H}{dB}$, as derived from a linear fit to each curve (some are not shown in Fig.~\ref{Hallmeas}a), as a function of $V_g $. The top x axis is the estimated total change in the electron number density, as explained in the caption of Fig.~\ref{Gmeas}. \label{Hallmeas}}
\end{figure}
Hall resistance and longitudinal resistance measurements as a function of back gate voltage were performed on the devices using conventional lock-in technique. Focusing first on the longitudinal resistance measurements, Fig.~\ref{Gmeas} presents the results obtained at a temperature of 4.2 K, normalized to the value at $V_g=0$. All the devices exhibited a strong increase in resistance when $V_g$ was changed toward negative values, however the response differed between devices. Note that in the narrow devices (width $<5 \mu m$) the change in the resistance was $20-40$ times of its original value. Measurements performed at 1.35 K, 10 K, 15 K and 21.5 K showed similar but weaker response at higher temperatures.
\par
  We assume that changing the gate voltage changes  the total carrier density in the 2DEG linearly, via the mutual capacitance between the gate and the 2DEG. This capacitance, as well as the initial carrier density at $V_g=0$, can be different for each device. As shown in the inset of Fig. 2a, the curves can be made to coincide by rescaling and shifting the x axis of the different curves. Thus we conclude that, in the narrower devices, the stronger response of the resistivity to back gate voltage can be explained by a different mutual capacitance between the gate and the 2DEG. The relative changes in fitted capacitance for different devices are presented in Fig.~\ref{Gmeas}b. Except for the narrowest bridge, the capacitance per unit area is  linear with the width of the Hall bar.
  
 We now show that the linear dependence is expected if the total capacitance is dominated by  the geometrical capacitance. In fact, in such devices the capacitance per unit area is  substantially different from that of an infinite planar capacitor, because the distance to the back gate is much greater than the typical size of the 2DEG.  This can be shown by solving an electrostatic model within the Thomas-Fermi approximation framework and assuming a non-zero density of states for the 2DEG near the Fermi surface. In addition, we show below that because of the properties of the STO/LAO interface, the smallest measured device is comparable in size to the Thomas-Fermi screening length. This causes the ``quantum capacitance'' (which is the finite quantum density of the states)  to dominate the total capacitance, and which may explain the experimental result. 
\par 
We considered a typical device geometry,  including the small Hall bar, the leads, the rectangular contacts and the back gate. The effect of the thin LAO layer was neglected due to the large difference in dielectric constant between LAO and STO. The potential of the back-gate was chosen to be zero, and therefore the chemical potential at the 2DEG is $-e\delta V_g$. At every point $\vec{r}=\left(x,y\right)$ of the 2DEG, this chemical potential obeys the  electrostatic equation
\begin{equation}
 -\delta V_g=\frac{\delta n(\vec{r})}{\nu}+ \frac1{2\pi\epsilon}\int d^{2}r'\left[\frac{\delta n\left(\vec{r'}\right)}{\left|\vec{r'}-\vec{r}\right|}-\frac{\delta n\left(\vec{r'}\right)}{\sqrt{\left(\vec{r'}-\vec{r}\right)^2+4d^2}}\right]
 \label{eq:mu0_1}
\end{equation}
  where $\delta n(\vec{r'})$ is the carrier density at point $\vec{r'}$ of the 2DEG, $\nu$ is the 2DEG density of states and $\epsilon\approx24000\epsilon_0$ is the dielectric constant of the  STO below the 2DEG. The last term describes the electrostatic potential, which is induced by the image charge due to the grounded back gate located at a distance $d=500\ \mu m$ below the 2DEG.  
  \par
 The size of the Hall bar in our samples is much smaller than that of the surrounding contacts and the distance to the back gate. In this case, the influence of the charges on the Hall bar on the charge distribution around it can be neglected. We thus separated the integral in Eq.~\eqref{eq:mu0_1}  into two parts: the contribution of the charges which belong to the Hall bar, say within some distance $r_0$ from it's center, and the contribution of all the charges at $r'>r_0$, namely those at the leads, at the contacts and at the back gate. The potential induced by the second group of charges on the first one depends linearly on the back gate voltage, and can be treated as an external potential  $-\alpha(\vec{r}) \delta V_g$, where $\alpha(\vec{r})$ is a unitless positive function. Consequently, Eq.~\eqref{eq:mu0_1} for $\vec{r}=0$ can be rewritten as 
\begin{equation}
\frac{\delta n(\vec{r}=0)}{\nu}+ \left(1-\alpha\right)\delta V_g +\frac{1}{2\pi\epsilon}\int_{r<r_{0}}d^{2}r'\frac{\delta n\left(\vec{r'}\right)}{\left|\vec{r'}\right|}=0\label{eq:mu0}.
\end{equation}
 where $\alpha\equiv\alpha(\vec{r}=0)$. The capacitance per area is given by
 
 \begin{equation}
 C_g\equiv - \frac{\delta n(\vec{r}=0)}{\delta V_g}= \left(1-\alpha\right)\frac{\epsilon}{l+l_{TF}},
 \end{equation}
 where $l=\frac{1} {2\pi n(\vec{r}=0)}\int_{r<r_{0}}d^{2}r'\frac{\delta n\left(\vec{r'}\right)}{\left|\vec{r'}\right|}$, and $l_{TF}=\frac{\epsilon}{\nu}$ is the 2D Thomas-Fermi screening length. 
 
 Note that $l$ is roughly related to the geometrical size of the Hall bar. For a Hall bar satisfying $l\gg l_{TF}$,  the capacitance per area is dominated by the geometrical capacitance, $C_g=\left(1-\alpha\right)\frac{\epsilon}{l}$. In the opposite case, where $l_{TF}\gg l$, the capacitance per area is dominated by the density of states, $C_g=\left(1-\alpha\right)\nu$.
 
Eq. ~\eqref{eq:mu0_1} was solved numerically for $\delta n(r)$ of the $10\,\mu m$ Hall bar with the contacts.  $\alpha$ was found to be 0.76, with less then $1\%$ sensitivity on the charges at the Hall-bar. For $l_{T
F}$ of few microns, We found $l\approx30 \mu m$ with a dependence of up to $10\%$ on $l_{T
F}$, resulting in a geometrical capacitance per area of $C_g\approx1.1\times10^{12}\  \frac {el}{cm^2V}$. This is more then four times larger than the naive calculation for a planar capacitor with d=500 nm resulting in $\frac{\epsilon}{ed}=2.7\times10^{11}\  \frac {el}{cm^2V}$.
      
   It should be emphasized that all of our devices had the same contacts geometry, leading to the same value of $\alpha$. In addition, the Hall bars were designed to have the same aspect ratio.  Therefore, $l$ is expected to be proportional to the width of the Hall bar, as indeed observed in Fig.~\ref{Gmeas}b. We conclude that in most of the devices the geometrical capacitance was dominant. Note that the values of the capacitance at the $3\ \mu m$ and $1.5\ \mu m$ wide devices are larger than the naive planar capacitor estimation by more than an order of magnitude. The deviation from the linear dependence of the capacitance in the $1.5\ \mu m$ wide device  may be due to the ``quantum capacitance'', $\nu$.
 
\par  
We now focus our discussion on the 3 $\mu m$ wide device. In Fig.~\ref{Hallmeas}a the results of Hall measurements for different values of back gate voltage are presented. For magnetic fields of up to $0.5\,T$, the Hall resistance depends linearly	on the magnetic field, at any gate voltage. The slope for each curve ${R_0}\equiv-\frac{dR_H}{dB}$ was derived by a linear fit. Fig.~\ref{Hallmeas}b presents the positive values of  $\left(eR_0\right)^{-1}$ as a function of the gate voltage. For a 2DEG system with one conduction band this parameter should be equal to the total electron number density, and thus it should increase simultaneously with back gate voltage. Suprisingly, we observed that $R_0^{-1}$ is not a monotonic function of the gate voltage. While in the voltage range of $-2V<V_g<0V$ $ R_0^{-1}$ behaved according to our expectation, for back gate voltages of $-7V<V_g<-2V$ an opposite trend was observed. The top x axis of Fig.~\ref{Hallmeas}b displays the change in the total carrier density for this range. This was obtained using the geometrical capacitance calculated theoretically for the $10\ \mu m$ Hall, multiplied by the measured ratio between the capacitances in the $3\ \mu m$ and the $10\  \mu m$  Hall bars (Fig. \ref{Gmeas}b). 
\par
Clearly, a 2DEG with a single band cannot produce these results. In order to gain more insight, we fit the data to a two-band model. In this model, the resistivity $\rho_{xx}$ and the Hall resistance $R_{xy}$  are  determined by four parameters: the carrier densities of the two bands, $n_1$ and $n_2$, and the corresponding mobilities, $\mu _1$ and $\mu _2$;
\begin{equation}
-R_H\approx R_{0}B=\frac{n_{1}+n_{2}\left(\frac{\mu_{2}}{\mu_{1}}\right)^{2}}{e\left(n_{1}+n_{2}\left(\frac{\mu_{2}}{\mu_{1}}\right)\right)^{2}}B,
\label{eq:R_0}\end{equation}
\begin{equation}
\rho_{xx}=\frac{1}{e\left(n_{1}\mu_{1}+n_{2}\mu_{2}\right)} . 
\label{eq:rho_xx}\end{equation}
For every gate voltage, given the values of the resistivity in Fig. \ref{Gmeas} and the Hall resistance in Fig. \ref{Hallmeas},  we have two out of the four parameters above which are left free. However, those parameters have some physical restrictions: first, both the mobilities and the densities are positive for each value of $V_g$. Second,  the change in the total number density, $\delta n_{tot}=\delta n_1 + \delta n_2$ is given by $e\delta n_{tot}=C_{g}\delta V_g$. Third, the ratio $\frac{\delta n_2}{\delta n_1}$ is constant and is proportional to the ratios of the density of states of the two bands. Finally, one would expect the mobilities of the two bands to decrease simultaneously with $V_g$. 
\par
 Under such restrictions, Eq.~\eqref{eq:R_0} was solved with densities satisfying $n_1+n_2>\left(eR_0\right)^{-1}>n_2$, and mobilities typically   satisfying  $\frac{\mu_2}{\mu_1}\approx 4-150$, depending on the initial densities, mass ratio and gate voltage. After finding a ratio $\frac{\mu_2}{\mu_1}$ that satisfies Eq.~\eqref{eq:R_0}, one can fit Eq.~\eqref{eq:rho_xx} to the curve of the $3 \ \mu m$ Hall bar in Fig.~\ref{Gmeas}a, to find the behavior of $\mu_1$ and $\mu_2$. The mobilities were indeed found to decrease with gate voltage for a wide range of mass ratios and initial densities. 
 \par
 In particular, one possible solution is obtained by choosing $n_2$ as the low carrier density values typically observed in SdH oscillations,  $n_2\sim4\times10^{12}$ $ \rm{cm}^{-2}$, [\onlinecite{Quantumos,SdHCaviglia}], together with $n_1\sim3\times10^{14}$ $\rm{cm}^{-2}$ as expected from the polar catastrophy \cite{Reinle-Schmitt_2012}. Fig. 4 shows the variation of the mobilities in both bands for the above initial carrier densities assuming a mass ratio of $m_1/m_2=0.1$. This scenario leads to both mobilities increasing by more then an order of magnitude, with a relative change of the two carrier densities of  less then 1/3 of a decade. Thus, the result within this scenario strongly supports a simultaneous metal-insulator phase transition in both bands. A different physical scenario is observed when starting out with the lower limit of $n_1$ under our restrictions: $n_1\sim4.7\times10^{13}$, and the corresponding value of $n_2\sim1\times10^{13}$. Using these starting conditions, we observe a phase transition only in a single band. 
 \begin{figure}
\includegraphics[ scale=0.40]{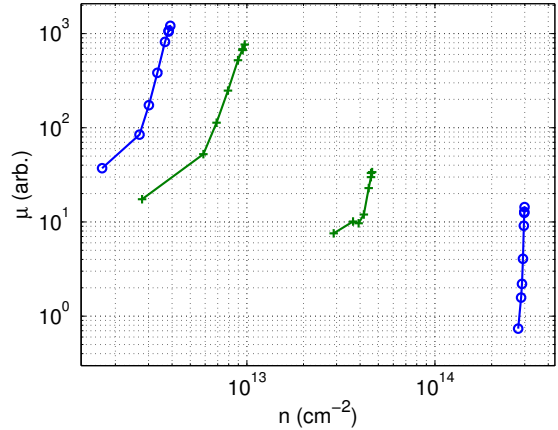}
\caption {(color online) Two possible solutions within the two band model which reproduce the observed longitudinal resistance in Fig.\ref{Gmeas} and Hall resistance in Fig.\ref{Hallmeas}. Blue circles markers correspond to initial conditions (at $V_g=0$) $n_1=3\times10^{14},\,n_2=4\times10^{12}$ and mass ratio $\frac{m_1}{m_2}=0.1$. Green plus markers correspond to initial conditions $n_1=4.7\times10^{13},\,n_2=1\times10^{13}$, and mass ratio $\frac{m_1}{m_2}=0.4$.  The lines are guides to the eye.\label{mu_vs_n}}
\end{figure} 
\par
In conclusion, this paper reports on a strong response of the transport properties of the 2DEG at LAO/STO interface upon applying several Volts of back gate voltage. The observation of the decrease of the Hall voltage during the depletion of the 2DEG excludes the possibility of a single conduction band. Moreover, The combined experimental data for the longitudinal and Hall resistance is consistent with the assumption of two conduction bands in which at least one band undergoes a metal to insulator transition. The large increase in longitudinal resistivity  depends  on the size of the mesa,  which strongly indicates differences in capacitance between the gate and the 2DEG for the different devices. An electrostatic model was used to calculate the geometrical capacitance between the back gate and the center of the small Hall bar, and found it to be much larger than the naive calculation for a plane capacitor.  The differences in the capacitances of  the various Hall bars are consistent with the theoretical prediction, except for the smallest device, in which quantum capacitance is expected to be more significant. 

The Hall measurements suggest the existence of a second conduction band with lower density and much higher mobility. This secondary band may be responsible for the unexpected period of SdH oscillations observed on similar devices\cite{Quantumos,SdHCaviglia}. The total carrier density, in turn, must be higher than the  values presented in Fig.~\ref{Hallmeas}b, namely higher than $3\times10^{13}\ cm^{-2}$. Further systematic study of transport properties as a function of back gate voltage and temperature is needed in order to deduce the exact nature of the observed transition. 
\begin{acknowledgments}
This research was supported by the Infrastructure program of the Israeli Ministry of Science and Technology under contract no. 01008080. I.N. acknowledges support from the Tel Aviv University Center for Nanoscience and Nanotechnology.

\end{acknowledgments}

\bibliographystyle{apsrev}
\bibliography{VWBIB}
\end{document}